\documentclass[12pt,draftcls,onecolumn]{IEEEtran}

\IEEEoverridecommandlockouts                              

\usepackage{amsfonts,amssymb,amsmath}
\usepackage[]{graphicx,graphics}
\usepackage{tikz}
\usetikzlibrary{automata}
\usepackage{color}
\usepackage{epstopdf}

\newtheorem{definition}{Definition}

\newtheorem{lemma}{Lemma}

\newtheorem{proposition}{Proposition}
\newtheorem{theorem}{Theorem}

\newtheorem{problem}{Problem}

\begin{document}

\title{Supervisor Synthesis to Thwart Cyber Attack with Bounded Sensor Reading Alterations}

\author{Rong Su\thanks{Rong Su is affiliated with Division of Control and Instrumentation, School of Electrical and Electronic Engineering, Nanyang Technological University, 50 Nanyang Avenue, Singapore 639798. Emails: rsu@ntu.edu.sg. The support from Singapore Ministry of Education Tier 1 Academic Research Grant M4011221.040 RG84/13 is gratefully acknowledged.}}

\maketitle

\begin{abstract}
One of the major challenges about cyber physical systems is how to prevent cyber attacks to ensure system integrity. There has been a large number of different types of attacks discussed in the modern control and computer science communities. In this paper we aim to investigate one special type of attacks in the discrete-event system framework, where an attacker can arbitrarily alter sensor readings after intercepting them from a target system in order to trick a given supervisor to issue control commands improperly, driving the system to an undesirable state. We first consider the cyber attack problem from an attacker point of view, and formulate an attack with bounded sensor reading alterations (ABSRA) problem. We then show that the supremal (or least restrictive) ABSRA exists and can be synthesized, as long as the plant model and the supervisor model are regular, i.e., representable by finite-state automata. Upon the synthesis of the supremal ABSRA, we present a synthesis algorithm, which ensures that a computed supervisor will be ABSRA-robust , i.e., either an ABSRA will be detectable or will not lead the system to an undesirable state.           
\end{abstract}


\section{Introduction\label{GD}}
A cyber-physical system (CPS) is a mechanism controlled or monitored by computer-based algorithms. Examples of CPS include smart grid, autonomous automobile systems, medical monitoring, process control systems, distributed robotics, and automatic pilot avionics, etc. The connection between the cyber part and the physical part heavily relies on communication networks, which has been raising a major security concern, as different types of cyber attacks can tamper the data collection processes and interfere safety critical decision making processes, which may cause irreparable damadges to the physical systems being controlled and to people who depend on those systems. \\

There has been a growing number of publications addressing the cyber security issues from both the computer science community, which focuses on the computer computation related issues, and the systems control community, which focuses on issues related to the system dynamics affected by cyber attacks. Recently, more and more efforts have been made in classifying different types of malicious attacks, assuming that the attackers are sufficiently intelligent \cite{CAS08} \cite{TPSJ12}, instead of merely just generating random failures, which is well studied in the fields of reliability and fault tolerant control. Typically, an intelligent attacker requires {\em system knowledge}, and abilities for {\em resource disclosure} and {\em resource disruption} in order to carry out a successful attack, which is covert to a system user until the attacker's goal of causing a damage to the system is achieved. So {\em covertness} and {\em damage infliction} are two major characteristics of a successful attack. By analyzing different intelligent cyber attacks, proper countermeasures may be developed to prevent a target system from being harmed by a specific type of attacks. \\

In this paper we study a special type of data deception attacks in the discrete-event system framework, where an attacker can intercept sensor measurements (or observations) modeled by observable events and alter them arbitrarily but with an upper bound imposed on the length of each altered observation sequence. By sending those altered observation sequences to a given supervisor, whose function is known to the attacker in advance, the attacker can deliberately and covertly guide the system to move into some undesirable states without making any change to the supervisor. The key challenge is how to ``fool'' the supervisor to make it believe that the system is operating correctly, while using the supervisor's own control functions to carry out the attack, i.e., to lead the system move into a bad state. To this end, we first propose the concept of {\em attack under bounded sensor reading alterations} (ABSRA), which can be modelled as a finite-state transducer, possessing the properties of covertness, damadge infliction and control feasibility under partial observations. Then we show that the supremal (or least restrictive) ABSRA exists and is computable via a specific synthesis algorithm, as long as both the plant model $G$ and the given supervisor $S$ are finitely representable. Upon this novel ABSRA synthesis algorithm, we present a supervisor synthesis algorithm, which can ensure that a nonempty synthesized supervisor will be ``robust'' to any ABSRA, in the sense that such an attack will either reveal itself to the supervisor due to abnormal system executions (so that proper contingent actions can be taken by the supervisor, which is nevertheless outside the scope of this paper) or will not be able to lead the system to a bad state (i.e., no damadge will be inflicted).\\

Our construction of an ABSRA model as a transducer is inspired by some recent work on opacity enforcement \cite{WL14}, which aims to use observable event insertions to prevent a potential attacker from correctly determining the actual state of a target system. Due to the different objectives of two works, the modeling details and synthesis algorithms  are completely different. There have been some works on cyber attack detection and prevention in the discrete-event community \cite{PSL11} \cite{TT06} \cite{CL15}, mainly from an adaptive fault tolerant control point of view, which heavily rely on real-time fault diagnosis to identify the existence of an attack and then take necessary robust or adaptive supervisory control actions. In those works the intelligence of an attacker is not considered, and an attack is treated as a fault. As a contrast, we do not rely on real time attack detection, but rely on prior knowledge of attack models, and simply build attack-robustness features into a supervisor to ensure that the supervisor will not be affected by any ABSRA unnoticeably. It is this  robust control nature distinguishes our works from existing DES-based cyber attack detection and prevention approaches, which fall in the adaptive control domain.\\               
    
The remainder of the paper is organized as follows. In Section II we review the basic concepts and operations of discrete event systems, and formulate an ABSRA synthesis problem, which is then solved in Section III, where we show that the supremal ABSRA exists and computable. In Section IV we present an algorithm to synthesize a supervisor, which is robust to any ABSRA. A simple yet realistic example runs through the entire paper to illustrate all relevant concepts and algorithms. Conclusions are drawn in Section V.  

\section{An ABSRA problem}
In this section we first recall some standard concepts used in the Ramadge-Wonham supervisory control paradigm. Then we introduce the concept of ABSRA, followed by a concrete ABSRA synthesis algorithm, which reveals that the supremal ABSRA is computable, as long as both the plant model and the given supervisor are regular. 

\subsection{Preliminaries on supervisory control}
Given an arbitrary finite alphabet $\Sigma$, let $\Sigma^*$ be the free monoid with the empty string $\epsilon$ being the unit element and the string concatenation being the monoid operation. Given two strings $s,t\in\Sigma^*$, $s$
is called a \emph{prefix substring} of $t$, written as $s\leq t$, if
there exists $u\in\Sigma^*$ such that $su=t$, where $su$ denotes
the concatenation of $s$ and $u$. Any subset $L\subseteq\Sigma^*$ is called a \emph{language}.  The \emph{prefix closure} of
$L$ is defined as $\overline{L}=\{s\in\Sigma^*|(\exists t\in L)\,
s\leq t\}\subseteq\Sigma^*$. Given
two languages $L,L'\subseteq\Sigma^*$, let
$LL':=\{ss'\in\Sigma^*|s\in L\,\wedge\, s'\in L'\}$ denote the
concatenation of two sets.  Let $\Sigma'\subseteq\Sigma$. A mapping $P:\Sigma^*\rightarrow\Sigma'^*$ is called the \emph{natural projection} with respect to $(\Sigma,\Sigma')$,
if
\begin{enumerate}
\item $P(\epsilon)=\epsilon$, 
\item $(\forall \sigma\in\Sigma)\, P(\sigma):=\left\{\begin{array}{ll} \sigma & \textrm{ if $\sigma\in\Sigma'$,}\\
    \epsilon & \textrm{ otherwise,}\end{array}\right.$ \item $(\forall s\sigma\in\Sigma^*)\, P(s\sigma)=P(s)P(\sigma)$.
\end{enumerate}
Given a language $L\subseteq\Sigma^*$, $P(L):=\{P(s)\in\Sigma'^*|s\in L\}$. The inverse image mapping of $P$ is
\[P^{-1}:2^{\Sigma'^*}\rightarrow 2^{\Sigma^*}:L\mapsto
P^{-1}(L):=\{s\in\Sigma^*|P(s)\in L\}.\] Given $L_1\subseteq\Sigma_1^*$ and $L_2\subseteq\Sigma_2^*$, the \emph{synchronous product} of $L_1$ and $L_2$
is defined as $L_1||L_2:=P_1^{-1}(L_1)\cap P_2^{-1}(L_2)$,
where $P_1:(\Sigma_1\cup\Sigma_2)^*\rightarrow\Sigma_1^*$ and $P_2:(\Sigma_1\cup\Sigma_2)^*\rightarrow\Sigma_2^*$ are natural projections. Clearly,
$||$ is commutative and associative.\\

A given target plant is modelled as a \emph{deterministic finite-state automaton}, $G=(X,\Sigma,\xi,x_0,X_m)$, where $X$ stands for the state set, $\Sigma$ for the
alphabet, $\xi:X\times\Sigma\rightarrow X$ for the (partial) transition function, $x_0$ for the initial state and $X_m\subseteq X$ for the
marker state set. We follow the notation system in \cite{Won07}, and use $\xi(x,\sigma)!$ to denote that the transition $\xi(x,\sigma)$ is defined. For each state $x\in X$, let $En_G(x):=\{\sigma\in\Sigma|\xi(x,\sigma)!\}$ be the set of events enabled at $x$ in $G$.  The domain of $\xi$ can be extended to $X\times\Sigma^*$, where $\xi(x,\epsilon)=x$ for all $x\in X$, and $\xi(x,s\sigma):=\xi(\xi(x,s),\sigma)$. The \emph{closed} behavior of $G$ is defined as $L(G):=\{s\in\Sigma^*|\xi(x_0,s)!\}$, and the \emph{marked} behavior of $G$ is $L_m(G):=\{s\in L(G)|\xi(x_0,s)\in X_m\}$. $G$ is \emph{nonblocking} if $\overline{L_m(G)}=L(G)$. We assume that the marker state set $X_m$ is partitioned into two disjoint sets $X_m=X_{d,m}\dot{\cup}X_{b,m}$, where $X_{d,m}$ is the set of \emph{desirable} states and $X_{b,m}$ denotes the set of \emph{bad} states.\\

We now recall the concept of supervisors. Let
$\Sigma=\Sigma_c\dot{\cup}\Sigma_{uc}=\Sigma_o\dot{\cup}\Sigma_{uo}$, where disjoint
$\Sigma_c$ ($\Sigma_o$) and $\Sigma_{uc}$ ($\Sigma_{uo}$) denote respectively
the sets of \emph{controllable} (\emph{observable}) and
\emph{uncontrollable} (\emph{unobservable}) events, respectively. Let $\Gamma:=\{\gamma\subseteq\Sigma|\Sigma_{uc}\subseteq\gamma\}$ be the collection of all \emph{control patterns}. A \emph{(feasible) supervisory control map of $G$ under partial observation $P_o:\Sigma^*\rightarrow\Sigma_o^*$} is defined as $V:L(G)\rightarrow\Gamma$, where 
\[(\forall s,s'\in L(G))\, P_o(s)=P_o(s')\Rightarrow V(s)=V(s').\]
For each $s\in L(G)$, $V(s)$ is interpreted as the set of events allowed to be fired after $s$. Thus, a supervisory control map will not disable any uncontrollable events, and will impose the same control pattern after strings, which cannot be distinguished based on observations.  Let $V/G$ denote the closed-loop system of $G$ under supervision of $V$, i.e.,
\begin{itemize}
\item $\epsilon\in L(V/G)$,
\item $L(V/G):=\{s\sigma\in L(G)|s\in L(V/G)\wedge \sigma\in V(s)\}$,
\item $L_m(V/G):=L_m(G)\cap L(V/G)$.
\end{itemize}
The control map $V$ is \emph{finitely representable} if $V/G$ can be denoted by a finite-state automaton, say $S=(Z,\Sigma,\delta,z_o,Z_m=Z)$ such that
\begin{itemize}
\item $L(S||G)=L(V/G)$ and $L_m(S||G)=L_m(V/G)$, where `$||$' is automaton product \cite{Won07},
\item $(\forall s\in L(S))\, E_S(s):=\{\sigma\in\Sigma|s\sigma\in L(S)\}=V(s)$,
\item $(\forall s,s'\in L(S))\, P_o(s)=P_o(s')\Rightarrow E_S(s)=E_S(s')$.
\end{itemize} 
It has been shown  that, as long as a closed-loop language $K\subseteq L_m(G)$ is \emph{controllable} \cite{RW87} and \emph{observable} \cite{LW88}, there always exists a finitely-representable supervisory control map $V$ such that $L_m(V/G)=K$ and $L(V/G)=\overline{K}$. From now on we assume that $V/G$ is finitely representable by $S$, which is called a \emph{supervisor}. We assume that $S$ is \emph{legal} in the sense that $L_m(S)\cap L_m(G)\subseteq \{s\in L_m(G)|\xi(x_0,s)\in X_{d,m}\}$, i.e., under the supervision of $S$, the plant $G$ should never enter any bad marker state.

\subsection{A sensor attack model}
We assume that an attacker can intersept each observable event generated by the plant $G$, and replace it by a sequence of observable events from $\Sigma_o$ in order to ``fool'' the given supervisor $S$, whose function is known to the attacker. Considering that in practice any event occurance takes an unnegligible amount of time, it is impossible for an attacker to insert an arbitrarily long observable sequence to replace a received observable event. For this reason, we assume that there exists a known natural number $n\in\mathbb{N}$ such that the length of any observable sequence that the attacker can insert is no more than $n$. Let $\Delta_n:=\{s\in\Sigma_o^*||s|\leq n\}$ be the set of all such bounded observable sequences, where $|s|$ denotes the length of $s$, and by convention, $|\epsilon|=0$. We model a sensor attack as a finite state transducer $A=(Y,\Sigma,\Delta_n,\eta,\theta,y_0,Y_m)$, where $Y$ is the state set, $\Sigma$ the input alphabet,  $\Delta_n$ the output alphabets, $y_0$ the initial state, $Y_m$ the marker state set, which is specifically set as $Y_m=Y$, and $\eta:Y\times \Sigma\times\Delta_n\rightarrow Y$ the (partial) transition map, where for all $\sigma\in\Sigma_{uo}$ and $y\in Y$, $\eta(y,\sigma,\epsilon)=y$, i.e., at each state $y$ all unobservable events are self-looped with $\epsilon$ as the output. This is natural because an attacker can only observe observable events, thus, will not make any move upon unobservable events. We still keep unobservable events here to make it easy for us for subsequent technical development. Clearly, $L(A)=L_m(A)\subseteq (\Sigma\times\Delta_n)^*$. Let $\psi:(\Sigma\times\Delta_n)^*\rightarrow \Sigma^*$ and $\theta:(\Sigma\times\Delta_n)^*\rightarrow \Delta_n^*$ be the \emph{input} and \emph{output} maps, respectively, where for each $\mu=(\sigma_1,u_1)(\sigma_2,u_2)\cdots (\sigma_l,u_l)\in (\Sigma\times\Delta_n)^*$,  $\psi(\mu)=\sigma_1\sigma_2\cdots \sigma_l$ and $\theta(s)=u_1u_2\cdots u_l$.\\

\begin{figure}[htb]
    \begin{center}
      \includegraphics[width=0.70\textwidth]{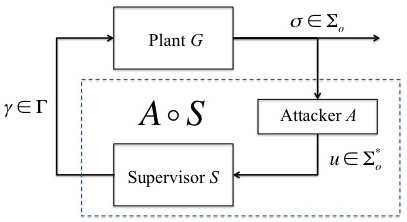}
    \end{center}
    \caption{The block diagram of an attack plan}
    \label{fig:Cyber-Security-0}
\end{figure}

The basic procedure of an attack is to intercept every single observable event $\sigma\in\Sigma_o$ generated by the plant $G$, replace it with some observable string $u\in\Delta_n$, and send $u$ to the supervisor $S$, in order to trick $S$ to issue a control command $\gamma\in\Gamma$ that may drive the plant $G$ towards a bad marker state. 
This attack procedure is depicted in Figure \ref{fig:Cyber-Security-0}.
The sequential composition of the attack $A$ and the supervisor $S$ essentially  forms a new supervisor, denoted as $A\circ S$, which receives an observable output $\sigma\in\Sigma_o$ and generates a control command $\gamma\in\Gamma$. The exact definition of this new supervisor reveals the nature of the attack, which is given below. The \emph{sequential composition} of $A$ and $S$ is a deterministic finite state transducer $A\circ S=(Y\times Z\cup \{d\},\Sigma\times\Delta_n,\zeta,(y_0,z_0),Y_m\times Z_m)$, where $d$ denotes the deadlocking dump state, and for each $(y,z), (y',z')\in Y\times Z$, $(\sigma,u)\in\Sigma\times\Delta_n$, $\zeta(y,z,\sigma,u)=(y',z')$ if one of the following holds,
\begin{itemize}
\item $\sigma\in\Sigma_{uo}\,\wedge\,  u=\epsilon\,\wedge\, \eta(y,\sigma,\epsilon)=y'\,\wedge\,\delta(z,\sigma)=z'$,
\item $\sigma\in\Sigma_o\,\wedge\, \eta(y,\sigma,u)=y'\,\wedge\, \delta(z,u)=z'$.
\end{itemize}
For each $(y,z)\in Y\times Z$, $(\sigma,u)\in\Sigma\times\Delta_n$, $\zeta(y,z,\sigma,u)=d$ if $\eta(y,\sigma,u)!$ but $\delta(z,u)$ is undefined. Thus, all transitions that go to the dumpt state $d$ may potentially reveal the attack, which, for an intelligent attack, should be avoided.\\ 

The \emph{impact} of $A$ on the closed-loop system $(G,S)$ is captured by the composition of the plant $G$ and the new supervisor $A\circ S$, i.e.,
\[G\times (A\circ S)=(X\times (Y\times Z\cup\{d\}), \Sigma\times\Delta_n,\kappa=\xi\times\zeta,(x_0,y_0,z_0),X_m\times Y\times Z),\] where for each $(x,w), (x',w')\in X\times (Y\times Z\cup\{d\})$, $(\sigma,u)\in\Sigma\times\Delta_n$, $(x',w')\in \kappa(x,w,\sigma,u)$ if $x'=\xi(x,\sigma)$ and $w'=\zeta(w,\sigma,u)$. Clearly, $G\times (A \circ S)$ is also a transducer, and it is not difficult to check that
\[G\times (A\circ S) = G\times (\textrm{Prefix}(G\times (A\circ S))\circ S),\]
where ``='' is in the sense of DES-isomorphism, and $\textrm{Prefix}(\cdot)$ denotes a function mapping one transducer to another transducer by simply marking every state. In other words, if $A$ is an attack model for the system $(G,S)$, then  $\hat{A}:=\textrm{Prefix}(G\times (A\circ S))$ is also an attack model, which has the same attack effect as that of $A$ on $(G,S)$. Since $\psi(L(\hat{A}))\subseteq L(G)$, we call $\hat{A}$ a \emph{canonical attack} with respect to $(G,S)$. Since for any attack, there exists a canonical attack, which has the same attack effect, from now on we only focus on canonical attacks. On the other hand, we will see that $A$ usually is stucturally simpler than its canonical one $\hat{A}$, whereas the latter is easier to compute. An interesting question is how to synthesize a simplified attack model $A$ from a given canonical attack model $\hat{A}$, which bears some similarity to the problem of supervisor reduction \cite{SW04}, and will be addressed in our future works.\\       

\begin{figure}[htb]
    \begin{center}
      \includegraphics[width=0.50\textwidth]{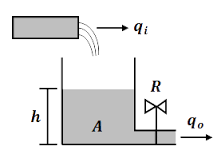}
    \end{center}
    \caption{A single tank system}
    \label{fig:Cyber-Security-1}
\end{figure}

\begin{figure}[htb]
    \begin{center}
      \includegraphics[width=0.90\textwidth]{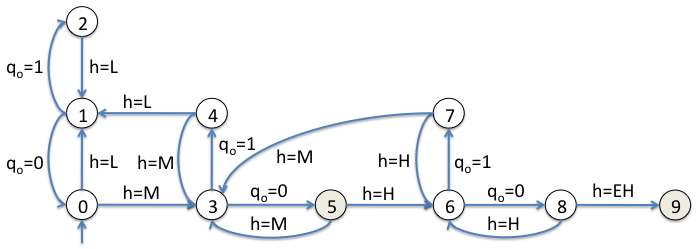}
    \end{center}
    \caption{Automaton model of the plant $G$}
    \label{fig:Cyber-Security-2}
\end{figure}

\begin{figure}[htb]
    \begin{center}
      \includegraphics[width=0.90\textwidth]{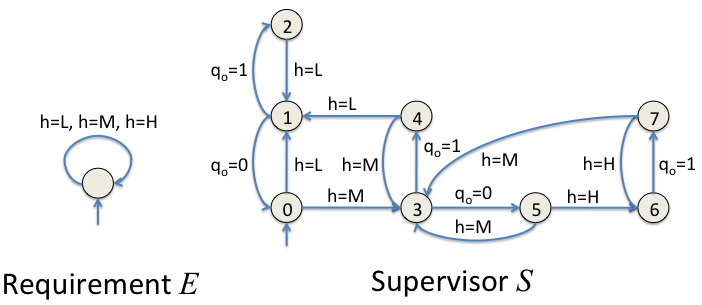}
    \end{center}
    \caption{Automaton models of a requirement $E$ (Left) and the supervisor $S$ (Right)}
    \label{fig:Cyber-Security-3}
\end{figure}

\begin{figure}[htb]
    \begin{center}
      \includegraphics[width=0.60\textwidth]{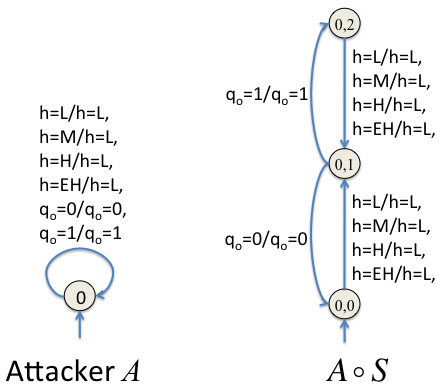}
    \end{center}
    \caption{Automaton models of an attack $A$ (Left) and the sequential composition $A\circ S$ (Right)}
    \label{fig:Cyber-Security-4}
\end{figure}

\begin{figure}[htb]
    \begin{center}
      \includegraphics[width=0.90\textwidth]{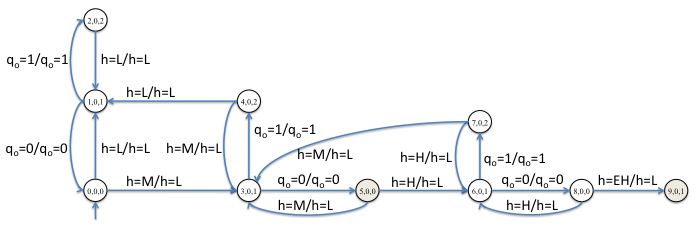}
    \end{center}
    \caption{Automaton models of $G\times (A\circ S)$}
    \label{fig:Cyber-Security-5}
\end{figure}

To illustrate the aforementioned concepts, let us go through a simple single-tank example depicted in Figure \ref{fig:Cyber-Security-1}, which consists of  
one water supply source whose supply rate is $q_i$, one tank, and one control valve at the bottom of the tank controlling the outgoing flow rate $q_o$, whose value depends on the valve opening and the water level $h$. We assume that the valve can only be fully open or fully closed to simplify our illustration, and in case of a full opening, the water level $h$ can only go down. The water level $h$ can be measured, whose value can trigger some predefined events, denoting the water levels: \emph{low} (h=L), \emph{medium} (h=M), \emph{high} (h=H), and \emph{extremely high} (h=EH).  We construct a simple discrete-event model of the system depicted in Figure \ref{fig:Cyber-Security-2},    
where the alphabet $\Sigma$ contains all events shown in the figure. All events are observable, i.e., $\Sigma_o=\Sigma$. Only the actions of opening the valve ($q_o=1$) and closing the valve ($q_o=0$) are controllable, and all water level events are uncontrollable. In the model we use a shaded oval to denote a marker state, i.e., state 5 and state 9 in Figure \ref{fig:Cyber-Security-2}. Assume that we do not want the water level to be extremely high, i.e., the event h=EH should not occur. Thus, state 9 is a bad marker state, i.e., $X_{d,m}=\{5\}$ and $X_{b,m}=\{9\}$. To prevent state 9 from being reached, we compose a requirement $E$ shown in Figure \ref{fig:Cyber-Security-3}, whose alphabet is $\{$h=L, h=M, h=H, h=EH$\}$, but the event h=EH is never allowed in the model. A supervisor $S$ can be synthesized by using the standard Ramadge-Wonham supervisory control paradigm, which is also depicted in Figure \ref{fig:Cyber-Security-3}.   
It is clear that the supervisor $S$ only opens the valve when the water level is high, i.e., it disables the event $q_o=0$ at state 6 when the event h=H occurs. Our intuition tells us that if an attack always change events of h=M, h=H, h=EH to the event h=L, then the supervisor will not prevent the water level from reaching the extreme high level, i.e., the event h=EH will happen. For this reason, we conjecture an attack model $A$ shown in Figure \ref{fig:Cyber-Security-4}, where water levels will be altered to h=L, whereas all other events will remain unchanged. 
The sequential composition $A\circ S$ indicates that, no matter which water level is reached, the attack $A$ always sends h=L to the supervisor $S$, which tricks it to believe that it is safe to allow the valve to be either closed or opened. The impact of $A$ on the closed-loop system $(G,S)$ is depicted in Figure \ref{fig:Cyber-Security-5}.
By marking every state in $G\times(A\circ S)$ we obtain a canonical attack model $\textrm{Prefix}(G\times(A\circ S))$.\\

\begin{proposition}\label{prop0}\textnormal{(1) $\theta(L_m(A\circ S))\subseteq L_m(S)$; (2) $L_m(A\circ S)\subseteq L_m(A)$; (3) $\psi(L_m(G\times (A\circ S)))=\psi(L_m(A\circ S))\cap L_m(G)$, $\psi(L(G\times (A\circ S)))=\psi(L(A\circ S))\cap L(G)$.\hfill $\Box$}\end{proposition}
Proof: By the above definition of sequential composition, the proposition follows.\hfill $\blacksquare$\\

\begin{proposition}\label{prop1}\textnormal{Given two attacks  $A_1$ and $A_2$ with the same input alphabet $\Sigma$ and output alphabet $\Delta_n$, assume that $L(A_1)\subseteq L(A_2)$. Then we have $L(A_1\circ S)\subseteq L(A_2\circ S)$.\hfill $\Box$}\end{proposition}
Proof: By the above definition of sequential composition, the proposition follows.\hfill $\blacksquare$\\

Given two attacks $A_1$ and $A_2$ with the same input alphabet $\Sigma$ and output alphabet $\Delta_n$, let $A_1\cup A_2$ be their union, which is a deterministic finite-state transducers. Then by the definition of transducer union, we have $L(A_1\cup A_2)=L(A_1)\cup L(A_2)$.\\

\begin{proposition}\label{prop2}\textnormal{$L((A_1\cup A_2)\circ S)=L(A_1\circ S)\cup L(A_2\circ S)$.\hfill $\Box$}\end{proposition} 
Proof: Since $L(A_1)\subseteq L(A_1\cup A_2)$ and $L(A_2)\subseteq L(A_1\cup A_2)$, by Prop. \ref{prop1} we have $L((A_1\cup A_2)\circ S)\supseteq L(A_1\circ S)\cup L(A_2\circ S)$. To show the other direction, for each string $\mu=(\sigma_1,u_1)\cdots (\sigma_n,u_n)\in L((A_1\cup A_2)\circ S)$, by the definition of the sequential composition, we know that $\mu\in L(A_1\cup A_2)$. Thus, either $\mu\in L(A_1)$ or $\mu\in L(A_2)$, which means either $\mu\in L(A_1\circ S)$ or $\mu\in L(A_2\circ S)$. Thus, $\mu\in L(A_1\circ S)\cup L(A_2\circ S)$, which concludes the proof.\hfill  $\blacksquare$\\

So far we have introduced a simple sensor attack model, and explained how this attack affects the closed-loop system. But we have not described what kind of sensor attacks can be considered intelligent. Next, we will introduce the concept of ABSRA.

\subsection{An ABSRA model}
Let $P_o:\Sigma^*\rightarrow\Sigma_o^*$ be the natural projection.  An intelligent canonical attack needs to possess the following properties:
\begin{enumerate}
\item Its insertions must be covert to the given supervisor, i.e.,
         \begin{equation}\theta(L(A))\subseteq L(S),\end{equation}
         namely the supervisor will not see any unexpected observable sequences from the attack. 
\item Any of its insertion sequence may potentally cause damages to $G$, i.e.,
        \begin{equation}\psi(L(G\times (A\circ S)))=\overline{\psi(L(A\circ S))\cap \{s\in L_m(G)|\xi(x_0,s)\in X_{b,m}\}},\end{equation}
    namely any sequence of insertions by the attack will cause $G$ to reach some bad state eventually. A weaker version of this property is described below:
        \[\psi(L(G\times (A\circ S)))\cap \{s\in L_m(G)|\xi(x_0,s)\in X_{b,m}\}\neq\varnothing,\]
which says that the attack $A$ will tamper the absolute correctness of the supervisor $S$ so that there exists some possibility that the system may reach some bad marker state.                
\item $A\circ S$ forms a standard supervisor for the plant $G$ that enforces normality \cite{LW88}, i.e.,
\begin{equation} P_o^{-1}(P_o(\psi(L(G\times(A\circ S)))))\cap L(G)=\psi(L(G\times (A\circ S))),\end{equation}
and 
\begin{equation}
(\forall \mu\in L(G\times (A\circ S)))\, \psi(En_{G\times (A_0\circ S)}(\kappa(x_0,y_0,z_0,\mu)))=En_S(\delta(z_0,\theta(\mu)),
\end{equation}
which denotes that at each state the attack will not intervene the event enablement by the supervisor because we consider only sensor attacks, not actuator attacks. 
\end{enumerate}
We call a nonempty model $A$ satisfying the aforementioned four properties (1)-(4)  an \emph{Attack with Bounded Sensor Reading Alterations} (ABSRA) of $(G,S)$.\\ 

It is not difficult to check that the attack $A$ shown in Figure \ref{fig:Cyber-Security-4} does not satisfy Property (1) because $S$ cannot fire $q_o=0$ before h=L, but $A$ can. Nevertheless, the sequential composition $A\circ S$ satisfies all three properties, thus, is an ABSRA. By the aforementioned discussions, we know that the canonical attack model $\textrm{Prefix}(G\times (A\circ S))$ is also an ABSRA.\\  

\begin{theorem}\label{thm1}\textnormal{Given a plant $G$ and a legal supervisor $S$, let $\{A_i|i\in I\}$ be a (possibly infinite) collection of ABSRA's with respect to $(G,S)$. Then $\cup_{i\in I}A_i$ satisfies properties (1)-(4).\hfill $\Box$}\end{theorem}
Proof: By Prop. \ref{prop2}, we know that $L((\cup_{i\in I}A_i)\circ S)=\cup_{i\in I}L(A_i\circ S)$. We now verify that $\cup_{i\in I}A_i$ satisfies all four properties.\\
(a) Since for each $i\in I$, $A_i$ is an ABSRA, we have that 
\[\theta(L(A_i))\subseteq L(S).\]
Thus, by Prop. \ref{prop2} we have that
\[\theta(L(\cup_{i\in I}A_i))\subseteq L(S).\]
(b) In addition, we have that for each $i\in I$,
        \[\psi(L(G\times (A_i\circ S)))=\overline{\psi(L(A_i\circ S))\cap \{s\in L_m(G)|\xi(x_0,s)\in X_{b,m}\}}.\]
Thus, by Prop. \ref{prop2} we have
        \[\psi(L(G\times ((\cup_{i\in I}A_i)\circ S)))=\overline{\psi(L((\cup_{i\in I}A_i)\circ S))\cap \{s\in L_m(G)|\xi(x_0,s)\in X_{b,m}\}}.\]
(c) Since for each $i\in I$, we have
\[ P_o^{-1}(P_o(\psi(L(G\times (A_i\circ S)))))\cap L(G)=\psi(L(G\times (A_i\circ S))),\]
we  get
\[ P_o^{-1}(P_o(\psi(L(G\times ((\cup_{i\in I}A)\circ S)))))\cap L(G)=\psi(L(G\times ((\cup_{i\in I}A_i)\circ S))).\]
The last property (4) can be easily checked. Thus, $\cup_{i\in I}A_i$ satisfies all four properties, and the theorem follows.\hfill $\blacksquare$\\

Theorem \ref{thm1} only implies that the least restrictive (or supremal) attack language exists. But it is not clear whether this supremal language is regular, i.e., whether it can be recognized by a finite-state transducer. Therefore, at this moment the existance of the supremal ABSRA is still unknown. We now state our main problem in this paper.\\

\begin{problem} Given a plant $G$ and a legal supervisor $S$, design an ABSRA $A$.\hfill $\Box$\\ \end{problem} 

In the next section we will show that the supremal attack language is regular, i.e., indeed the supremal ABSRA exists, and is computable.

\section{Synthesis of an ABSRA}

We first recall the concepts of controllability \cite{RW87}, and normality \cite{LW88}. Because we deal with both finite-state automata and finite-state transducers, to make notations simple, we introduce a general purpose alphabet $\Lambda$, which can be either $\Lambda=\Sigma$ or $\Lambda= \Sigma\times\Delta_n$, depending on a specific application context. Let $\Lambda_{uc}\subseteq\Lambda$ and $\Lambda_o\subseteq\Lambda$ be an uncontrollable alphabet and an observable alphabet respectively, where if $\Lambda=\Sigma\times\Delta_n$ then $\Lambda_o:=\Sigma_o\times\Delta_n$.  Let $\hat{P}_o:\Lambda^*\rightarrow\Lambda_o^*$ be the natural projection. In case that $\Lambda=\Sigma\times\Delta_n$, we have  $\hat{P}_o(\epsilon)=\epsilon$, $\hat{P}_o(\sigma,u)=(\sigma,u)$ if $\sigma\in\Sigma_o$, or $\epsilon$ otherwise, and $\hat{P}_o(\mu (\sigma,u))=\hat{P}_o(\mu)\hat{P}_o(\sigma,u)$. When we mention a finite-state transitional structure $\mathcal{G}$, we mean that $\mathcal{G}$ is either a finite-state automaton or a finite-state transducer.\\

\begin{definition}\label{Def1}\textnormal{Given a finite-state transitional structure $\mathcal{G}$, a sublanguage $K\subseteq L_m(\mathcal{G})$ is \emph{controllable} w.r.t. $\mathcal{G}$ and $\Lambda_{uc}$, if $\overline{K}\Lambda_{uc}\cap L(\mathcal{G})\subseteq\overline{K}$.\hfill $\Box$\\}
\end{definition}

\begin{definition}\label{Def2}\textnormal{Given a finite-state transitional structure $\mathcal{G}$,  a sublanguage $K\subseteq L_m(\mathcal{G})$ is \emph{normal} w.r.t. $\mathcal{G}$ and $\Lambda_o$, if $\hat{P}_o^{-1}(\hat{P}_o(K))\cap L(\mathcal{G})= K$.\hfill $\Box$\\}
\end{definition}

Given a finite-state transitional structure $\mathcal{G}$, whose alphabet is $\Lambda$, and a requirement $\mathcal{E}\subseteq\Lambda^*$, let \[\mathcal{CN}(\mathcal{G},\mathcal{E}):=\{K\subseteq L_m(\mathcal{G})\cap\mathcal{E}|K\textrm{ is controllable w.r.t. $\mathcal{G}$ and $\Lambda_{uc}$ }\wedge\, \overline{K}\textrm{  is normal w.r.t. $\mathcal{G}$ and $\Lambda_o$}\}.\] By an argument similar to the one used in \cite{RW87}, we can derive that the supremal controllable and normal sublanguage of $L_m(\mathcal{G})$ exists, denoted as $\textrm{sup}\mathcal{CN}(\mathcal{G},\mathcal{E})$, such that for all $K\in\mathcal{CN}(\mathcal{G},\mathcal{E})$, we have $K\subseteq\textrm{sup}\mathcal{CN}(\mathcal{G},\mathcal{E})\in\mathcal{CN}(\mathcal{G},\mathcal{E})$.\\

In our setup, an attack is able to arbitrarily alter an observable event. Thus, each event $(\sigma,u)\in\Sigma\times\Delta_n$ is considered controllable, as the attack can choose not to use this alteration. Under this consideration, the uncontrollable alphabet $\Lambda_{uc}$ is actually empty. Thus, in the following attack model synthesis, we do not explicitly require controllability. This may sound a bit unusual because we do have an uncontrollable alphabet $\Sigma_{uc}$ for the plant $G$ - how those uncontrollable events affect the attack model synthesis? If we carefully check the properties of an ABSRA, we can see that Property (4) actually implicitly enforces controllability with respect to $\Sigma_{uc}$ because it requires the attack not to change the event enablement of the supervisor $S$ at the current state, and since by default the supervisor $S$ ensures controllability with respect to $\Sigma_{uc}$, and so does the attack model.\\ 

Assume that there exists $\Sigma_{o,p}\subseteq\Sigma_o$, which denotes a set of \emph{protected} observable events that cannot be altered by an ABSRA, i.e., given an attack model $A=(Y,\Sigma,\Delta_n,\eta,\theta,y_0,Y_m)$, for all $y\in Y$, $(\sigma,u)\in\Sigma_{o,p}\times\Delta_n$, we have that $\eta(y,\sigma,u)!\Rightarrow u=\sigma$. We now undertake the following ABSRA synthesis  procedure.\\
\textbf{Procedure 1: (ABSRA Synthesis )}
\begin{enumerate}
\item Input: a plant $G=(X,\Sigma,\xi,x_0,X_m)$, a supervisor $S=(Z,\Sigma,\delta,z_0,Z)$ and $\Sigma_{o,p}$.
\item Construct a single-state transducer $A_0=(Y,\Sigma,\Delta_n,\eta,\theta,y_0,Y)$, where $Y=\{y_0\}$ and the transition map $\eta$ encodes transitions labeled by a subset of $(\Sigma_o\setminus\Sigma_{o,p})\times\Delta_n\cup \Sigma_{u,p}\times\Sigma_{u,p}\cup \Sigma_{uo}\times\{\epsilon\}$, denoting all observable event alterations that the attack wants to consider. 
\item Let $\mathcal{E}_0:=\{\mu\in L_m(G\times (A_0\circ S))|\xi(x_0,\psi(\mu))\in X_{b,m}\}$ be a requirement. 
\item Undertake the following iteration on $k=1,\cdots$
\begin{enumerate}
\item Compute $K_k:=\textrm{sup}\mathcal{CN}(G\times (A_0\circ S),\mathcal{E}_{k-1})$. 
\item  Check property (4) in the definition of ABSRA. If it holds, then go to Step 5). Otherwise, set $\mathcal{E}_k:=\{\mu\in K_k|\psi(En_{G\times (A_0\circ S)}(\kappa(x_0,y_0,z_0,\mu)))=En_S(\delta(z_0,\theta(\mu)))\}$ and continue the iteration on $k$.
\end{enumerate} 
\item Output: $A_*$, which recognizes $\overline{K_{k+1}}$.\hfill $\Box$\\
\end{enumerate}

\begin{lemma}\textnormal{Procedure 1 terminates finitely.\hfill $\Box$}\end{lemma}
Proof: Assume that $\mathcal{E}_0$ is recognized by a transducer $R_0$, whose state set is $W_0$. Then $K_1$ is recognizable by a transducer, say $R_1$, whose state set is a subset of $X\times Y\times Z\times W_0$. It is not difficult to check that for all $\mu, \mu'\in K_1$, if they hit the same state in $R_1$, then we know that
\[\psi(En_{G\times (A_0\circ S)}(\kappa(x_0,y_0,z_0,\mu)))=En_S(\delta(z_0,\theta(\mu)))\]
if and only if 
\[\psi(En_{G\times (A_0\circ S)}(\kappa(x_0,y_0,z_0,\mu')))=En_S(\delta(z_0,\theta(\mu'))).\]
In other words, $\mu\in \mathcal{E}_1$ if and only if $\mu'\in\mathcal{E}_1$. Thus, for each state in $R_1$, either all strings hitting that state are in $\mathcal{E}_1$ or none of them are in $\mathcal{E}_1$, namely $\mathcal{E}_1$ is recognized by a sub-transducer $\hat{R}_1$ of $R_1$. Suppose the state set of $\hat{R}_1$ is $W_1\subseteq X\times Y\times Z\times W_0$. By the property of automaton composition, we know that there exists a transducer $R_2$ recognizing $K_2$ such that the state set of $R_2$ is a subset of $X\times Y\times Z\times W_1$. Since $W_1\subseteq X\times Y\times Z\times W_0$, we know that $R_2$ is DES-isomorphic to a sub-transducer of $R_1$. By using the same argument, we can check that each $K_k$ is recognized by a transducer, which is DES-isomorphic to a sub-transducer of $R_1$. In addition, the state sets of those sub-transducers form a monotonic non-increasing sequence with respect to set inclusion. Thus, in a finite number of iterations, a fixed sub-transducer will be reached, whose language is $K_k$. This means Procedure 1 must terminate finitely.\hfill $\blacksquare$\\

\begin{lemma}\label{lem1}\textnormal{Let $K_k$ and $A_*$ be computed in Procedure 1. Then $K_k:=\textrm{sup}\mathcal{CN}(G\times (A_*\circ S),\mathcal{E}_{k-1})=L_m(G\times (A_*\circ S))$.\hfill $\Box$}\end{lemma}
Proof: By the proof of Lemma 1 we know that $A_*$ is DES-isomorphic to the prefix closure of a sub-transducer of $G\times (A_0\circ S)\times R_0$. Then by the definitions of sequential composition and transducer product,  the lemma follows.
\hfill $\blacksquare$\\ 

\begin{theorem}\textnormal{$A_*$ obtained in Procedure 1 is the supremal ABSRA of $(G,S)$. \hfill $\Box$}\end{theorem}
Proof: (a) We first show that $A_*$ is an ABSRA, i.e., $A_*$ satisfies properties (1)-(4). It is clear that when the algorithm terminates, property (4) must hold. So we only focus on properties (1)-(3). By the definition of $A_*$ and Prop. 1, we know that \[\theta(L(A_*))=\theta(\overline{\textrm{sup}\mathcal{CN}(G\times (A_*\circ S),\mathcal{E}_{k-1})})\subseteq L(S).\] 
By Lemma 2, we have $K_k=L_m(G\times (A_*\circ S))$. Since $K_k\subseteq\mathcal{E}_{k-1}$ and $\psi(\mathcal{E}_{k-1})\subseteq \{s\in L_m(G)|\xi(x_0,s)\in X_{b,m}\}$, by Prop. 1 we have 
\begin{eqnarray*}\overline{\psi(K_k)}=\overline{\psi(L_m(A_*\circ S))\cap L_m(G)} &= & \overline{ \psi(L(A_*\circ S))\cap \{s\in L_m(G)|\xi(x_0,s)\in X_{b,m}\}}.\end{eqnarray*} 

For the third property, by Lemma 2 we know that $K_k:=\textrm{sup}\mathcal{CN}(G\times (A_*\circ S),\mathcal{E}_{k-1})$. Thus, by the definition of normality and the nonblocking property associated with a sub-transducer of $G\times (A_0\circ S)\times R_0$, which recognizes $K_k$, we have
\[P_o^{-1}(P_o(\psi(L(G\times (A_*\circ S)))))\cap L(G)=\psi(L(G\times (A_*\circ S))).\]
This concludes our proof that $A_*$ is an ABSRA.\\
(b) Next, we show that $A_*$ is the supremal ABSRA. Let $\hat{A}$ be an ABSRA of the system. Clearly, $L(\hat{A})\subseteq L(A_0)$. Due to the controllability of $S$ and the assumption that $\hat{A}$ is an ABSRA, i.e., it must satisfy property (4), it is easy to check that $L_m(G\times (\hat{A}\circ S))$ is controllable w.r.t. $L_m(G\times (A_0\circ S))$ and $\Sigma_{uc}$. Since $\hat{A}$ must satisfy Property (3), we know that $L_m(G\times (\hat{A}\circ S))$ must be normal w.r.t. $L_m(G\times (A_0\circ S))$ and $\Sigma_o$. In addition, $\hat{A}$ satisfies property (4). Thus, we can easily detive that $L_m(G\times (\hat{A}\circ S))\in \mathcal{CN}(G\times (A_0\circ S),\mathcal{E}_{k-1})$, namely, $L_m(G\times (\hat{A}\circ S))\subseteq K_k=\textrm{sup}\mathcal{CN}(G\times (A_0\circ S),\mathcal{E}_{k-1})$. This means $L(\hat{A})\subseteq L(A_*)$, which concludes the proof of the theorem. \hfill $\blacksquare$\\ 

\begin{figure}[htb]
    \begin{center}
      \includegraphics[width=0.90\textwidth]{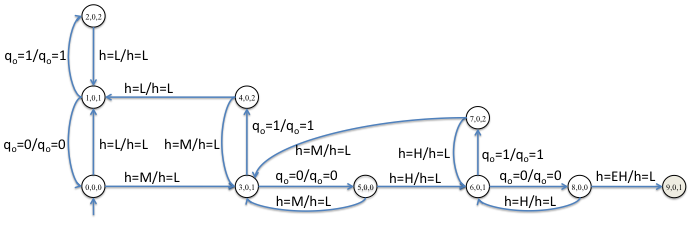}
    \end{center}
    \caption{A transducer recognizing $K_*$}
    \label{fig:Cyber-Security-6}
\end{figure}

As an illustration, we apply Procedure 1 to the plant $G$ shown in Figure \ref{fig:Cyber-Security-2} and the supervisor $S$ shown in Figure \ref{fig:Cyber-Security-3}. We can see that the sensor attack model $A$ in Figure \ref{fig:Cyber-Security-4} is actually $A_0$ in Procedure 1 because all events in the model are observable. The composition $A_0\circ S=A\circ S$ is shown in Figure \ref{fig:Cyber-Security-4}. The outcome of $G\times (A_0\circ S)$ is shown in Figure \ref{fig:Cyber-Security-5},  
which is isomorphic to $G$. This is not surprising because any string in $L(G)$ may be potentially extended to the bad marker state. The requirement $\mathcal{E}$ is simply the same as $G\times (A_0\circ S)$. Clearly, we know that   $K_k$ is recognizable by a transducer shown in Figure \ref{fig:Cyber-Security-6}, which is almost the same as $G\times (A_0\circ S)$, except that the only marker state is that bad marker state due to the requirement $\mathcal{E}$.   
Since all events are observable, from $\hat{P}_o(L(A_*))=\hat{P}_o(\overline{K_*})$ we can derive that $L(A_*)=\overline{K_*}$. Thus, $A_*$ can be chosen by marking every state in $G\times (A_0\circ S)$, i.e., $A_*=\textrm{Prefix}(G\times (A_0\circ S))$, which means $A_*$ is actually a canonical attack of $A_0$ with respect to $(G,S)$. By Theorem 2 we know that $A_*$ is the supremal ABSRA of $(G,S)$.


\section{Synthesis of an ABSRA-robust supervisor}
In the previous section we discuss how to design an ABSRA model to interrupt a given system's operations from an attacker's point of view. In this section we present a synthesis approach to design a supervisor, which is ``robust'' to any ABSRA in the sense that either the attack is not covert or incurs no damage to the system. \\

Recall that an ABSRA affects a target system $(G,S)$ by altering the sequence of observable events, which tricks $S$ to issue commands improperly. By protecting observable events from being altered unnoticeably can in principle effectively deter an ABSRA. An observable event in this framework denotes a specific set of strongly associated measurements. For example, in the aforementioned single-tank system, the event h=H may either be associated with one simple water level measurement or possibly several sensor measurements such as the actual water level, and the corresponding pressure on the bottom of the tank - the more sensor measurements associated with the event, the harder for an attack to alter the event without being detected. When applying suitable encryption techniques, it is even more complicated for an attack to complete the job. Thus, it is indeed technically feasible to prevent observable events from being altered by either adopting new secure information transmission technologies or introducing more sensors to significantly increase the complication of altering the corresponding observable event without being detected. Nevertheless, there is always a financial consideration. An attractive solution to a potential industrial user is to identify only critical observable events, which, when being protected from external alterations, will lead to a supervisor robust to any ABSRA. \\

\begin{problem}\textnormal{Given a plant $G$, a requirement $E$, and a protected observable alphabet $\Sigma_{o,p}\subseteq\Sigma_o$, synthesize a supervisor $S$ such that there is no ARSRA of the closed-loop system $(G,S)$.\hfill $\Box$\\ }\end{problem}

With the same notations used in the previous section, let $\mathcal{CN}(G,E)$ be the collection of all controllable and normal supervisors \cite{Won07}. Let $S_0=\textrm{sup}\mathcal{CN}(G,E)$, which always exists and computable, as long as $E\subseteq\Sigma^*$ is regular. Our goal is to design a supervisor $S\in \mathcal{CN}(G,E)$ such that Procedure 1 returns an ampty ABSRA $A_*$ with respect to the given protected observable alphabet $\Sigma_{o,p}$. To this end, we present the following synthesis procedure:\\
\textbf{Procedure 2: (ABSRA-Robust Supervisor Synthesis)}
\begin{enumerate}
\item Input: a plant $G$, a requirement $E$ and a protected observable alphabet $\Sigma_{o,p}$.
\item Compute $\hat{K}=\textrm{sup}\mathcal{CN}(G,E)$. If $\hat{K}=\varnothing$, terminate. Otherwise, assume $\hat{K}$ is recognized by a finite-state automaton $\hat{S}$, and continue.
\item Compute $A_*$ by using Procedure 1, i.e., computer $K_k=\textrm{sup}\mathcal{CN}(G\times (A_0\circ \hat{S}),\mathcal{E}_k)$. 
\item Compute $K:=\textrm{sup}\mathcal{CN}(G,L(\hat{S})-\theta(K_k))$.
\item Output: a recognizer $S$ of $\overline{K}$.\hfill $\Box$\\
\end{enumerate}

\begin{theorem}\label{thm3}\textnormal{Given a plant $G$, a requirement $E\subseteq\Sigma^*$, and a protected observable alphabet $\Sigma_{o,p}$, let $S$ be computed above. If $L_m(S)\neq\varnothing$, then we have
$\textrm{sup}\mathcal{CN}(G\times (A_0\circ S),\mathcal{E}_k)=\varnothing$, where $\mathcal{E}_k$ is defined in Procedure 1, i.e., there is no ABSRA $A$ of $(G,S)$.\hfill $\Box$
}\end{theorem}  
Proof: Assume that it is not true. Then there exists an ABSRA  $A$ such that $L(A)=\overline{\tilde{K}_k}$, where $\tilde{K}_k=\textrm{sup}\mathcal{CN}(G\times (A_0\circ S),\mathcal{E}_k)\neq\varnothing$. Since $S$ is controllable and normal with respect to $G$, and $L(S)\subseteq L(\hat{S})$, we can get that $\tilde{K}\in  \mathcal{CN}(G\times (A_0\circ \hat{S}),\mathcal{E}_k)$. Since $K_k=\textrm{sup}\mathcal{CN}(G\times (A_0\circ \hat{S}),\mathcal{E}_k)$, we know that $\tilde{K}\subseteq K_k$. But on the other hand, we know that $\theta(\tilde{K})\subseteq K\subseteq L(S)-\theta(K_k)$, i.e., $\tilde{K}\nsubseteq K_k$, which leads to a contradiction. Thus, the ABSRA $A$ does not exist. \hfill $\blacksquare$\\

We would like to emphasize here again that, although Theorem \ref{thm3} indicates that there is no ABSRA $A$ for the closed-loop system $(G,\hat{S})$, it does not mean that a sensor reading alteration attack will not be carried out by an attacker. But
such an attack will either reveal itself to the supervisor before it achieves its attack goal due to abnormal system executions (so that proper contingent actions such as system shutdown can be taken by the supervisor, which is outside the scope of this paper) or will not be able to lead the system to a bad state. \\

In Theorem \ref{thm3}, if $K=\varnothing$, then with the given protected observable alphabet $\Sigma_{o,p}$, there does not exist a supervisor $S$ that is ABSRA-robust. We face the following synthesis problem.\\

\begin{problem}\textnormal{Given a plant $G$ and a requirement $E$, compute a protected observable alphabet $\Sigma_{o,p}\subseteq\Sigma_o$ of the minimum size, which allows a nonempty ABSRA-robust supervisor $S$ to exist.\hfill $\Box$\\ }\end{problem}   

It is clear that Problem 3 is solvable in the sense that it is decidable whether there exists such a $\Sigma_{o,p}$ with the minimum size because we can simply enumerate each subset $\Sigma_{o,p}\subseteq \Sigma_o$, and apply Procedure 2 on $\Sigma_{o,p}$  to compute the corresponding supervisor $S$. Since there is a finite number of such subsets, this brutal-force method will terminate, and provide a protected observable alphabet of the minimum size together with the corresponding supervisor, if it exists. The computational complexity of this procedure is certainly high, which is exponential to $|\Sigma_o|$, but polynomial to the sizes of $G$ and $E$ due to our adoption of normality to handle observability.  If the size of $\Sigma_o$ is big, to find a computationally viable algorithm that can solve Problem 3 becomes important, which will be addressed in our future works.\\

\begin{figure}[htb]
    \begin{center}
      \includegraphics[width=0.90\textwidth]{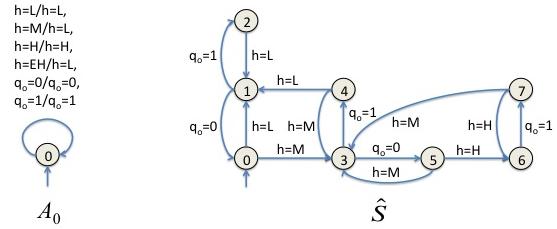}
    \end{center}
    \caption{Models of $A_0$ and $\hat{S}$}
    \label{fig:Cyber-Security-7}
\end{figure}

\begin{figure}[htb]
    \begin{center}
      \includegraphics[width=1.0\textwidth]{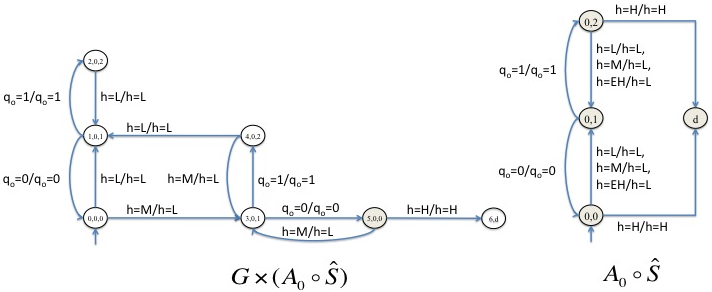}
    \end{center}
    \caption{Models of $A_0\circ \hat{S}$ (Right) and $G\times (A_0\circ \hat{S})$ (Left)}
    \label{fig:Cyber-Security-8}
\end{figure}

We now use that simple single-tank system to illustrate how to use Procedure 2 to compute an ABSRA-robust supervisor, and how to determine a minimum protected observable alphabet, which allows the existence of an ABSRA-robust supervisor. Let $\Sigma_{o,p}=\{$h=H$\}$. The model of $A_0$ and $\hat{S}$ are shown in Figure \ref{fig:Cyber-Security-7}.  
When we run Procedure 1, we first compute $G\times (A_0\circ \hat{S})$. The outcome is depicted in Figure \ref{fig:Cyber-Security-8}.
We can see that $G\times (A_0\circ \hat{S})$ contains no bad marker state in $X_{b,m}$. Thus, in Procedure 1 we have $\mathcal{E}_0=\varnothing$, which returns $K_1=\textrm{sup}\mathcal{CN}(L_m(G\times(A_0\circ \hat{S})),\mathcal{E}_0)=\varnothing$. After that, in Step 4) of Procedure 2, we have that $K=\textrm{sup}\mathcal{CN}(L_m(G),L_m(\hat{S}))=L_m(\hat{S})$. Thus, $\hat{S}$ is an ABSRA-robust supervisor for $G$ with respect to the given $\Sigma_{o,p}$. Clearly, it is a solution to Problem 3 because we cannot find any other protected observable alphabet with a size smaller than 1, which can render an ABSRA-robust supervisor.     


\section{Conclusions}
In this paper we have first introduced the concept of ABSRA, upon which we have shown that the supremal ABSRA exists and computable, as long as the plant model $G$ and the supervisor $S$ are finitely representable, i.e., their languages are regular. After that, we have brought in the problem of synthesizing an ABSRA-robust supervisor, and shows that it is possible to find a minimum protected observable sub-alphabet, which may render an ABSRA-robust supervisor.\\ 

It is interesting to point out that, if we replace the third property of an ABSRA model with a weaker observability property, e.g., the standard observability \cite{LW88}, the supremal ABSRA may not exist any more. Nevertheless, the existence of an ABSRA is still decidable and computable (with possibly a higher computational complexity), as this ABSRA synthesis problem is equivalent to a synthesis problem of centralized supervisory control under partial observation, which has been shown solvable \cite{YL06}.  Fortunately, the normality property can be easily satisfied in reality, as it only requires that only observable and controllable events can be disabled in online applications - in real industrial applications, it is typical that all control commands are observable. For this reason, the supervisor synthesis approach proposed in this paper aiming to defy ABSRA is practically feasible.

\noindent\textbf{Acknowledgement}

The idea of this paper was originated from a discussion between the author and Prof Stephane Lafortune on opacity enforcement. Without such an inspiring discussion, this paper would never be formed. For this reason, the author would like to thank Prof Lafortune for his contribution.


\begin{thebibliography}{10}








\bibitem{CL08}
C. Cassandra and S. Lafortune. {\em Introduction to Discrete Event Systems} (2nd Ed.), Springer, 2008.








\bibitem{LW88}
F. Lin and W. M. Wonham.\newblock On observability of discrete-event
systems.\newblock {\em Information Sciences}, 44(3):173-198, 1988.





\bibitem{Pap94}
C. H. Papadimitriou. {\em Computational Complexity}. Addison Wesley, 1994.

\bibitem{RW87} P.J. Ramadge and W.M. Wonham.
\newblock Supervisory control of a class of discrete event systems.
\newblock {\em SIAM J. Control and Optimization}, 25(1):206--230, 1987.








\bibitem{Su13}
R. Su. Discrete-event modeling of multi-agent systems with broadcasting-based parallel composition. {\em Automatica}, 49(11):3502-3506, 2013.

\bibitem{SSR10}
R. Su, J.H. van Schuppen and J.E. Rooda.
\newblock Aggregative synthesis of distributed supervisors based on automaton
abstraction. \newblock {\em IEEE Trans. Automatic Control}, 55(7):1627-1640, 2010.

\bibitem{SSR12}
R. Su, J.H. van Schuppen, J.E. Rooda. \newblock Maximally permissive coordinated distributed supervisory control of nondeterministic discrete-event systems. {\em Automatica}, 48(7):1237-1247, 2012.

\bibitem{SW04}
R. Su, W. M. Wonham. \newblock Supervisor reduction for discrete-event systems. {\em Journal of Discrete Event Dynamic Systems}, 14(1):31-53, 2004.



\bibitem{YL06} T.S. Yoo and S. Lafortune.
\newblock Solvability of centralized supervisory control under partial observation..
\newblock {\em Discrete Event Dynamic Systems: Theory and Applications}, 16(4):527--553, 2006.


\bibitem{Won07} W.~M. Wonham.
\newblock {\em Supervisory Control of Discrete-Event Systems}.
\newblock Systems Control Group, Dept. of ECE, University of Toronto. URL:
  www.control.utoronto.ca/DES, 2014.

\bibitem{WR87} W.M. Wonham and P.J. Ramadge.
\newblock On the supremal controllable sublanguage of a given language.
\newblock {\em SIAM J. Control and Optimization}, 25(3):637--659, 1987.

\bibitem{CAS08}
A. A. Cardenas, S. Amin, and S. Sastry. Secure control: towards survivable
cyber-physical systems. In {\em Proc. 28th International Conference
on Distributed Computing Systems Workshops}, 2008, pp. 495–500.

\bibitem{TPSJ12}
A. Teixeira, D. Perez, H. Sandberg, and K. H. Johansson. Attack models and scenarios for networked control systems. In {\em Proc. 1st International Conference on High Confidence Networked Systems}, 2012, pp. 55–64.

\bibitem{WL14}
Y. C. Wu and S. Lafortune. Synthesis of insertion functions for enforcement of opacity security properties. {\em Automatica}, 50(5):1336-1348, 2014.

\bibitem{PSL11} 
A. Paoli, M. Sartini and S. Lafortune. Active fault tolerant control of
discrete event systems using online diagnostics. {\em Automatica}, 47(4):639 – 649, 2011.

\bibitem{TT06}
D. Thorsley and D. Teneketzis. Intrusion detection in controlled
discrete event systems. In {\em Proc. 45th IEEE Conference on Decision
and Control}, 2006, pp. 6047–6054.

\bibitem{CL15}
L. K. Carvalho, Y. Wu, R. Kwong and S. Lafortune. Detection and prevention of actuator enablement attacks in supervisory control systems. In {\em Proc. 13th International Workshop on Discrete Event Systems}, 2016, pp. 298–305.

\end{thebibliography}
\end{document}